\documentclass[12pt]{revtex4}
\usepackage{epsfig}

\begin{document}

\title{Perturbative Expansion of Irreversible Work in Fokker-Planck Equation {\it \`{a} la} Quantum Mechanics}

\author{T. Koide}
\email{tomoikoide@gmail.com,koide@if.ufrj.br}
\affiliation{Instituto de F\'{\i}sica, Universidade Federal do Rio de Janeiro, C.P.
68528, 21941-972, Rio de Janeiro, Brazil}

\begin{abstract}
We discuss the systematic expansion of the solution of the Fokker-Planck equation 
with the help of the eigenfunctions of the time-dependent Fokker-Planck operator. 
The expansion parameter is the time derivative of the external parameter which controls the form of an external potential.
Our expansion corresponds to the perturbative calculation of the adiabatic motion in quantum mechanics. 
With this method, we derive a new formula to calculate the irreversible work order by order, 
which is expressed as the expectation value with a pseudo density matrix.
Applying this method to the case of the harmonic potential, we show that the first order term of the expansion gives the exact result.
Because we do not need to solve the coupled differential
equations of moments, our method simplifies the calculations of various 
functions such as the fluctuation of the irreversible work per unit time.
We further investigate the exact optimized protocol to minimize the irreversible work by calculating its variation 
with respect to the control parameter itself.
\end{abstract}

\maketitle

\section{Introduction}

The accelerating development of experimental methods enables us to 
access individual thermal random processes at microscopic scales.
Because 
the typical scale of the system is very small and then thermal fluctuations 
play important roles, thermodynamics is not directly applicable to describe this system.
The establishment of the coarse-grained description of the small-fluctuating systems is 
an intriguing subject of statistical physics and nanophysics.

Such a system is often modeled by the Brownian motion \cite{hanngi-review,broeck-review,sekimoto-book,sei-rev}.
Then the work which is exerted or absorbed by the system of the Brownian particle is represented 
as the change of the form of the external confinement potential.
In practice, any protocol of this change is realized within a finite time period and thus we cannot 
avoid the loss of energy.
However, we can still consider the optimized protocol which minimizes the irreversible work for a fixed time period 
\cite{sekimoto-book,sekimoto-sasa,schmiedl-seifert2007,schmiedl-seifert2008,gomez,koning,aurell,geiger,then,dechant,bona,zulkow,crook,all}.
The investigation of the optimized protocol is important to construct more efficient nanomachines.

As is well-known, the distribution function of the Brownian particle is described by the Fokker-Planck equation and 
the irreversible work is calculated from it.
In principle, the behavior can be solved numerically,  
but such an approach will not be promising for investigating the optimization problem.
On the other hand, when the work is expressed as an analytic function of the control parameter, 
we can apply the variational scheme to find the optimized protocol \cite{sekimoto-sasa,sekimoto-book}. 
In fact, the optimization has been exclusively studied for the harmonic potential which can be solved exactly 
\cite{schmiedl-seifert2007,schmiedl-seifert2008,gomez,speck,mazon,raybov,imparato,cohen,kwon,hol2}. 
\footnote{The optimization in the logarithmic-harmonic potential is studied in Ref.\ \cite{hol}.}

In an exact calculation, the macroscopic quantities of the Brownian particle is expressed 
as the functions of the moments of the position of the Brownian particle \cite{schmiedl-seifert2007,schmiedl-seifert2008,gomez,speck}. 
Therefore, to calculate the fluctuations of, for example, the irreversible work, we need to solve 
highly coupled differential equations of the moments. 
In addition, all methods which have been employed to solve the exactly solvable models 
are not easily applicable to systems which have more general potentials \cite{blickle,dieterich}.
To study the optimized protocol in more general situations, 
we thus need to develop a systematic method to find an analytic expression of the irreversible work.

In this work, we develop a perturbative expansion method to calculate the solution of the Fokker-Planck equation.
The expansion basis is the eigenfunctions of the time-dependent Fokker-Planck operator, and the expansion parameter 
is the time derivative of the external parameter which controls the form of the external potential.
Our expansion corresponds to the perturbative calculation of the adiabatic motion in quantum mechanics. 
Then all results are expressed in integral forms and we do not need to solve differential equations of the moments. 
This feature is an advantage of our approach compared to the moment method.
To confirm the consistency of the expansion, we apply this method to the harmonic potential and derive the 
irreversible work and its fluctuation.
We then show that the first and the second order correction terms of the expansion are sufficient to reproduce the exact results.
We further calculate the variation of the irreversible work with respect to the control parameter itself, not to the moment (variance), 
and derive the exact equation to determine the optimized protocol. 
Because of the complex integro-differential equation, it is difficult to find the exact solution. 
Instead we discuss the approximated solution of the integro-differential equation.

This paper is organized as follows.
In Sec.\ \ref{sec:bi}, we derive the bi-orthogonal system with the eigenfunctions of the time-dependent Fokker-Planck operator.
Using this, we develop the perturbative expansion of the solution of the Fokker-Planck equation in Sec.\ \ref{sec:adi}, and 
derive the expansion formula to calculate the irreversible work in Sec.\ \ref{sec:work}.
In Sec.\ \ref{sec:harmonic}, we apply the obtained result to the harmonic potential and investigate the exact the exact irreversible work, 
its fluctuation and the optimized control parameter.
Section \ref{sec:conc} is devoted to the concluding remarks.

\section{Bi-orthogonal system} \label{sec:bi}

In this section, we discuss the definition of the expansion basis.
A similar expansion basis is discussed, for example, in Ref.\ \cite{gardiner}.
In this section, we generalize the method to the case of 
the time-dependent Fokker-Planck operator \cite{namiki}. 
Note that the eigenvalue theory of the time-periodic Fokker-Planck operator (the Kolmogorov operator) 
is discussed in Ref.\ \cite{caceres} and the properties found below are consistent with the result.

We consider a Brownian particle which is confined in an external confinement potential $V$ and interacts with a thermal bath with a fixed temperature $T$.
When the inertial term is negligible and the system is spatially one dimensional, 
the distribution function of the particle $\rho(x,t)$ is described by the Fokker-Planck equation,
\begin{equation}
\partial_t \rho(x,t) = \left[ \frac{1}{\nu\beta} \partial^2_x 
+ \frac{1}{\nu}\partial_x V^{(1)}(x,a_t) \right] \rho(x,t) \equiv {\cal L}_t (x)\rho(x,t), \label{fpeq}
\end{equation}
where $\nu$ is the constant friction coefficient and $\beta = 1/(k_B T)$ with $k_B$ being the Boltzmann constant.
Note that $V^{(n)} = \partial^n_x V$ with $V$ being the potential energy, and $a_t$ is a time dependent external parameter which controls the form of the potential.
Differently from Ref.\ \cite{caceres}, $a_t$ is an arbitrary function of time.
In the following, we consider the case where the particle distribution function $\rho(x,t)$ vanishes quickly at infinite distance, 
${\displaystyle \lim_{x\rightarrow \pm \infty}} \rho (x,t) = 0$.

The eigenvalue and eigenfunction of the time-dependent Fokker-Planck operator ${\cal L}_t(x)$ are defined by 
\begin{eqnarray}
{\cal L}_t (x) \rho_n (x,a_t) &=& -\bar{\lambda}_n (t) \rho_n(x,a_t) \label{eigen-L}.
\end{eqnarray}
Note that ${\cal L}_t(x)$ is not self-adjoint and the eigenfunctions do not form a complete set in general. 
However, we can further introduce the eigenfunctions defined by 
\begin{eqnarray}
{\cal L}^\dagger_t (x) \tilde{\rho}_n (x,a_t) = -\bar{\lambda}_n (t) \tilde{\rho}_n (x,a_t), 
\end{eqnarray}
where 
\begin{eqnarray}
{\cal L}^\dagger_t (x) = \left[ \frac{1}{\nu \beta} \partial^2_x - \frac{1}{\nu}V^{(1)}(x,a_t) \partial_x  \right]. 
\nonumber
\end{eqnarray}
Then $\rho_n(x,a_t)$ and $\tilde{\rho}_n(x,a_t)$ form a bi-orthogonal system as is shown soon later and the solution of the Fokker-Planck equation can be 
expanded by these eigenfunctions.

To construct $\rho_n(x,a_t)$ and $\tilde{\rho}_n(x,a_t)$, we should note that both of ${\cal L}_t$ and ${\cal L}^\dagger_t$ 
are characterized by ${\cal H}_t$, which is defined by 
\begin{equation}
e^{G(x,a_t)/2} {\cal L}_t (x) e^{-G(x,a_t)/2} = e^{-G(x,a_t)/2} {\cal L}^\dagger_t (x) e^{G(x,a_t)/2}  
= - {\cal H}_t, \label{lld-relation}
\end{equation}
where
\begin{eqnarray}
&& {\cal H}_t  = -\frac{1}{\nu\beta} \partial^2_x - \frac{1}{2\nu} V^{(2)}(x,a_t) + \frac{\beta}{4\nu} (V^{(1)}(x,a_t))^2, \nonumber \\
&& G(x,a_t) = \beta V(x,a_t). \nonumber
\end{eqnarray}
This quantity ${\cal H}_t$ can be regarded as the Hamiltonian operator which has the potential given by 
$- \frac{1}{2\nu} V^{(2)}(x,a_t) + \frac{\beta}{4\nu} (V^{(1)}(x,a_t))^2$. 
Then we introduce the eigenfunctions of ${\cal H}_t $ as 
\begin{equation}
{\cal H}_t  u_n (x,a_t) = \lambda_n(t) u_n(x,a_t),
\end{equation}
which form a complete orthonormal set, 
\begin{eqnarray}
\int dx u_n(x,a_t) u_m (x,a_t) &=& \delta_{n,m}, \\
\sum_n  u_n(x,a_t) u_n(x',a_t) &=& \delta(x-x').
\end{eqnarray}
We should further notice that because there is no degeneracy for the eigenvalue of the one-dimensional Hamiltonian 
($\lambda_n \neq \lambda_m$ for $n\neq m$),  
the eigenfunction $u_n$ is given by a real function ($u^*_n = u_n$).

Once the eigenvalues and eigenfunctions of ${\cal H}_t$ are found, owing to Eq.\ (\ref{lld-relation}), 
the eigenfunctions of the time-dependent Fokker-Planck operators are given by 
\begin{eqnarray}
\rho_n (x,a_t) &=& e^{-G(x,a_t)/2} u_n (x,a_t), \label{rho-u1}\\
\tilde{\rho}_n (x,a_t) &=& u_n (x,a_t)e^{G(x,a_t)/2}, \label{rho-u2}
\end{eqnarray}
and the corresponding eigenvalue is given by 
\begin{equation}
\bar{\lambda}_n (t)= \lambda_n (t).
\end{equation}
One can easily confirm that these eigenfunctions form the following bi-orthogonal system, 
\begin{eqnarray}
\int dx \tilde{\rho}_n (x,a_t) \rho_m (x,a_t) &=& \delta_{n,m}, \\
\sum_n \tilde{\rho}_n (x,a_t) \rho_n (x',a_t) &=& \delta(x-x').
\end{eqnarray}

We can further show the following properties as the universal natures of the present eigenvalue problem.
\begin{enumerate}

  \item The smallest eigenvalue is given by zero and other eigenvalues are larger than zero. Because there is no degeneracy for the eigenvalues, we can set, 
\begin{equation}
0 = \bar{\lambda}_0 (t)< \bar{\lambda}_1 (t)< \bar{\lambda}_2 (t)< \cdots,
\end{equation}
without loss of generality.

  \item The eigenfunctions $\rho_0 (x,a_t)$ and $\tilde{\rho}_0 (x,a_t)$ for $\bar{\lambda}_0(t)$ are given by 
\begin{eqnarray}
\rho_0 (x,a_t) &=& \frac{1}{\sqrt{Z(a_t)}}e^{- \beta V(x,a_t)}, \\
\tilde{\rho}_0 (x,a_t) &=& \frac{1}{\sqrt{Z(a_t)}}, \label{tilrho0}
\end{eqnarray}
respectively. Here we introduced 
\begin{equation}
Z(a_t) = \int dx e^{-\beta V(x,a_t)}.
\end{equation}

\end{enumerate}

\vspace{2cm}

These properties can be shown as follows.
Note that the Hamiltonian operator can be reexpressed as 
\begin{eqnarray}
{\cal H}_t = \frac{1}{\nu\beta} B^\dagger B,  \nonumber 
\end{eqnarray}
where
\begin{eqnarray}
B &=& \partial_x + \frac{\beta}{2}V^{(1)}(x,a_t), \nonumber \\
B^\dagger &=& - \partial_x + \frac{\beta}{2}V^{(1)}(x,a_t). \nonumber 
\end{eqnarray}
Therefore the eigenvalues are more than or equal to zero, 
\begin{eqnarray}
\bar{\lambda}_n (t)= \lambda_n (t)= \frac{1}{\nu \beta} \int dx |B u_n(x,a_t)|^2 \ge 0. \label{eqn:eigenv}
\end{eqnarray}
From the absence of the degeneracy of the eigenfunctions, $\bar{\lambda}_n(t) \neq \bar{\lambda}_m(t)$ for $n \neq m$.
Moreover, the Fokker-Planck equation has a stationary state, which can be interpreted as the 
eigenfunction of zero eigenvalue. 
Thus, without loss of generality, we can set as
$0 = \bar{\lambda}_0 (t)<\bar{\lambda}_1 (t)< \bar{\lambda}_2 (t)< \cdots $.

From Eq.\ (\ref{eqn:eigenv}), it is seen that the normalizable eigenfunction for $\bar{\lambda}_0 (t)$ is 
given by the solution of $B u_0 (x,a_t) = 0$.
However, for the case of $V^{(2)}(x,a_t) > 0$, we can discuss as follows.
The operator $B$ satisfies 
\begin{eqnarray}
[{\cal H}_t, B] = - \frac{1}{\nu}V^{(2)}(x,a_t) B. \nonumber 
\end{eqnarray}
Therefore, we can show  
\begin{eqnarray}
{\cal H}_t (Bu_0(x,a_t)) = \left( \lambda_0 (t) - \frac{1}{\nu}V^{(2)} (x,a_t) \right) (Bu_0(x,a_t)). \nonumber 
\end{eqnarray}
Because $u_0 (x,a_t)$ has the lowest eigenvalue of ${\cal H}$, 
it should be the solution of the equation $Bu_0 (x,a_t) = 0$, leading to 
\begin{eqnarray}
u_0 (x,a_t) = \frac{1}{\sqrt{Z(a_t)}}e^{- \beta V(x,a_t)/2}, \nonumber
\end{eqnarray}
Then $\rho_0(x,a_t)$ and $\tilde{\rho}_0(x,a_t)$ are calculated using Eqs.\ (\ref{rho-u1}) and (\ref{rho-u2}).

There is an important remark. 
The above operators $B$ and $B^{\dagger}$, generally do not correspond to the lowering and raising operators in quantum mechanics. 
In fact, we can show that 
\begin{equation}
[{\cal H}_t, B^\dagger] = \frac{1}{\nu}B^\dagger V^{(2)} (x,a_t), \label{prob-raising}
\end{equation}
and $B^\dagger V^{(2)} (x,a_t) \neq V^{(2)}(x,a_t)  B^\dagger $.
Therefore we cannot obtain excited states by multiplying $B^\dagger$ to $u_0 (x,a_t)$.
It is however not the case with the harmonic potential as is seen in Sec.\ \ref{sec:harmo}.

For the sake of simplicity, we introduce the following bra-ket notation as quantum mechanics,  
\begin{eqnarray}
\rho_n (x,a_t) &=& \langle x | n,a_t \rangle, \nonumber \\
\tilde{\rho}_n (x,a_t) &=& \langle n,a_t | x \rangle, \nonumber \\
\langle x | \hat{\cal L}_t | n,a_t \rangle &=& 
{\cal L}_t (x)
\rho(x,t), \nonumber \\ 
\langle n,a_t  | \hat{\cal L}_t | x \rangle &=& 
{\cal L}^\dagger_t (x)
\tilde{\rho}(x,t), \nonumber \\ 
\int dx | x \rangle \langle x | &=& 1.\nonumber 
\end{eqnarray}
Here $| x \rangle$ ($\langle x |$) is an eigenfunction of the position operator $\hat{x}$, $\hat{x}| x \rangle = x | x \rangle$ 
($\langle x | \hat{x}  = \langle x | x$).
Other operators are introduced as quantities satisfying $\langle x | \hat{A} | x' \rangle = A(x) \delta (x-x')$, for example,  
\begin{eqnarray}
\langle x | \hat{G}(a_t) | x' \rangle &=& G(x,a_t) \delta (x-x'), \nonumber \\
\langle x | \hat{V}(a_t) | x' \rangle &=& V(x,a_t) \delta (x-x'), \nonumber \\
\langle x | \hat{\cal L}_t | x' \rangle &=& {\cal L}_t (x) \delta (x-x'), \nonumber \\ 
\langle x | 1 | x' \rangle &=& \delta (x-x'). \nonumber  
\end{eqnarray}
Then the Fokker-Planck equation is symbolically expressed as 
\begin{equation}
\partial_t | \rho(t) \rangle = \hat{\cal L}_t | \rho(t)\rangle, \label{eqn:fpeq-general}
\end{equation}
and the conditions satisfied for the eigenfunctions are summarized as  
\begin{eqnarray}
\langle n, a_t | m, a_t \rangle &=& \delta_{n,m}, \label{eqn:norma}\\
\sum_n | n, a_t \rangle \langle n, a_t | &=& 1.
\end{eqnarray}

In the following, we use this bi-orthogonal set to expand the solution of the Fokker-Planck equation.
We should note the difference between our approach and quantum mechanics: 
the bra-ket vectors represent the probability amplitude in quantum mechanics, while we use it to expand 
the probability density. In addition, a bra (ket) vector is normalized by calculating the scalar product with the corresponding ket (bra) vector as is shown 
in Eq.\ (\ref{eqn:norma}), 
but the normalization of the probability density is determined by $\int dx \rho(x,t) = 1$.

\section{Adiabatic expansion} \label{sec:adi}

There exist various expansion methods to find the solution of the Fokker-Planck equation. 
See Refs.\ \cite{hesam,dreger,talkner,welles2014} and references therein.
In our method, we expand the solution of the Fokker-Planck equation 
in terms of the eigenfunctions of the time-dependent Fokker-Planck operator which are obtained in the previous section.
Expanding $| \rho(t) \rangle$ in terms of $|n,a_t \rangle$, we find 
\begin{equation}
| \rho (t) \rangle = \sum_{n=0} C_n (t) e^{- \theta_n(t) - \int^t_{t_i} ds \bar{\lambda}_n (s) }| n, a_t \rangle,
\end{equation}
where
\begin{eqnarray}
C_n (t) &=& \langle n,a_t | \rho(t) \rangle e^{\theta_n(t) + \int^t_{t_i} ds \bar{\lambda}_n (s) }, \\
\theta_n(t) &=& \int^t_{t_i} ds \langle n, a_s | n, \dot{a}_s \rangle.
\end{eqnarray}
Here $t_i$ is an initial time and we introduced the notation 
\begin{equation}
| n, \dot{a}_t \rangle = \partial_t | n, {a}_t \rangle.
\end{equation}

Substituting this into the Fokker-Planck equation, we find the equation for the coefficient,
\begin{eqnarray}
\partial_t C_n (t) 
= - \sum_{m \neq n} C_m(t) e^{\int^s_{t_i} d\tau (\bar{\lambda}_n (\tau) - \bar{\lambda}_m (\tau))}  e^{\theta_n (s) - \theta_m (s)}  
 \langle m,a_t | n, \dot{a}_t \rangle. \nonumber 
\end{eqnarray}
On the other hand, operating the time derivative to Eq.\ (\ref{eigen-L}), we find
\begin{eqnarray}
\langle m,a_t | \dot{\hat{\cal L}}_t | n, a_t \rangle = -\dot{\bar{\lambda}}_n (t)\delta_{m,n} 
+ (\bar{\lambda}_m (t) - \bar{\lambda}_n (t)) \langle m, a_t | n, \dot{a}_t \rangle. \nonumber
\end{eqnarray}
Solving this, for $m\neq n$, $\langle m,a_t | n, \dot{a}_t \rangle$ is reexpressed as 
\begin{eqnarray}
\langle m,a_t | n, \dot{a}_t \rangle = \frac{\langle m,a_t | \dot{\hat{\cal L}}_t | n, a_t \rangle}{\bar{\lambda}_m (t) - \bar{\lambda}_n (t)}. \nonumber
\end{eqnarray}
Using this expression, we obtain the following differential equation 
\begin{eqnarray}
\partial_t {C}_n 
=
- \sum_{n\neq m} {C}_m (t) e^{\int^t_{t_i} ds (\bar{\lambda}_n (s) - \bar{\lambda}_m (s))}  e^{\theta_n (t) - \theta_m (t)}  
\frac{\langle n, a_t |\dot{\hat{\cal L}}_t | m, a_t \rangle}{\bar{\lambda}_n (t) - \bar{\lambda}_m (t)} \label{diffeq},
\end{eqnarray}
which can be formally solved as  
\begin{eqnarray}
{C}_n (t) 
=
{C}_n (t_i) 
- \int^t_{t_i} ds \sum_{n\neq m} {C}_m (s) e^{\int^s_{t_i} d\tau (\bar{\lambda}_n (\tau) - \bar{\lambda}_m (\tau))}  e^{\theta_n (s) - \theta_m (s)}  
\frac{\langle n, a_s |\dot{\hat{\cal L}}_s | m, a_s \rangle}{\bar{\lambda}_n (s) - \bar{\lambda}_m (s)}. \label{eq:cn}
\end{eqnarray}
From this equation, we can determine the coefficient iteratively.

This is similar to the derivation of the wave function in the adiabatic motion in quantum mechanics, and then $\theta_n (t)$ 
can be identified with the quantity corresponding to Berry's geometrical phase \cite{berry}.

It should be noted that the coefficient $C_0(t)$ is not affected by the 
second term on the right hand side of Eq.\ (\ref{eq:cn}), because 
\begin{eqnarray}
\langle 0, a_t | \dot{\hat{\cal L}}_t | m, a_t \rangle 
= \frac{1}{\nu \sqrt{Z(a_t)}} \int dx\ \partial_x \{ \dot{V}^{(1)}(x,a_t)\ \rho_m (x,a_t) \} = 0.\nonumber
\end{eqnarray}
Here we used that $\langle 0, a_t | x \rangle$ is constant due to Eq.\ (\ref{tilrho0}).
Therefore $C_0(t)$ is given by the initial condition itself, 
\begin{equation}
{C}_0 (t) = {C}_0 (t_i). \label{cond-c0}
\end{equation}
Moreover, $\theta_0 (t)$ is calculated as 
\begin{equation}
\theta_0 (t) = \int^t_{t_i} ds \langle 0, a_s | 0, \dot{a}_s \rangle = \ln \sqrt{\frac{Z(a_t)}{Z(a_i)}},
\end{equation}
with $a_i \equiv a_{t_i}$.

\subsection{Evolution from equilibrium state}

In the following, we limit our discussion to the evolution from the equilibrium state defined by 
\begin{eqnarray}
\rho_{eq} (x,t_i) = \langle x | \rho_{eq} (t_i) \rangle = \frac{1}{Z(a_i)} e^{-\beta V(x,a_i)}. \nonumber
\end{eqnarray}
Then the initial coefficient is given by 
\begin{eqnarray}
{C}_n (t_i) = \delta_{n,0} \langle 0, a_i | \rho_{eq} (t_i) \rangle 
= \delta_{n,0} \langle 0, a_i | x \rangle
= \delta_{n,0} \frac{1}{\sqrt{Z(a_i)}}. \nonumber 
\end{eqnarray}
In this case, the expansion of $| \rho (t) \rangle$ is simplified as 
\begin{equation}
| \rho (t) \rangle = \langle 0, a_i | \rho_{eq} (t_i) \rangle \sum_n D_n (t) e^{- \theta_n(t) 
- \int^t_{t_i} ds \bar{\lambda}_n (s) } | n, a_t \rangle,
\end{equation}
where the new coefficient $D_n(t)$ is determined by the following equation,
\begin{eqnarray}
D_n (t) 
&=& 
\delta_{n,0} - \int^t_{t_i}  ds \sum_{m\neq n} D_m (s) 
e^{\int^t_{t_i} ds (\bar{\lambda}_n (s)-\bar{\lambda}_m (s)) } 
e^{\theta_n(t) - \theta_m(t)} 
\frac{\langle n, a_s | (\partial_a \hat{\cal L}_s) | m, a_s \rangle}{\bar{\lambda}_n (s) 
- \bar{\lambda}_m (s)} \dot{a}_s \label{dn}.
\end{eqnarray}
Note that, because of Eq.\ (\ref{cond-c0}), 
\begin{equation}
D_0 (t) = 1. \label{d0}
\end{equation}
Then the expectation values can be expressed symbolically as 
\begin{eqnarray}
\langle A \rangle  
&=& 
\int dx \rho (x,t) A(x) \nonumber \\
&=&
\int dx A(x) 
\langle 0, a_i | \rho_{eq} (t_i) \rangle \sum_n D_n (t) e^{- \theta_n(t) - \int^t_{t_i} ds \bar{\lambda}_n (s) } \rho_n (a_t)
\nonumber \\
&=&
\int dx \sum_n D_n (t) e^{- \theta_n(t) - \int^t_{t_i} ds \bar{\lambda}_n (s) }
\langle 0, a_i | x \rangle A(x)   \langle x | n, a_t \rangle 
\nonumber \\
&\equiv& 
{\rm Tr} [A \hat{\rho}(t)],
\end{eqnarray}
where we introduced the pseudo density matrix defined by 
\begin{equation}
\hat{\rho} (t)= 
\sum_n D_n (t) e^{- \theta_n(t) - \int^t_{t_i} ds \bar{\lambda}_n (s) } | n, a_t \rangle\langle 0, a_i |.
\end{equation}
In this derivation, Eq.\ (\ref{tilrho0}) is used.

One can easily confirm that the pseudo density matrix satisfies 
\begin{eqnarray}
1 &=& {\rm Tr \hat{\rho}(t)}, \label{trace}\\
\hat{\rho}^2(t) &=& \hat{\rho}(t). \label{2=1}
\end{eqnarray}
See also Eq.\ (\ref{eqn:change}).
Moreover the diagonal component is given by the solution of the Fokker-Planck equation, 
\begin{eqnarray}
\langle x | \hat{\rho}(t) | x \rangle = \langle x | \rho(t) \rangle = \rho (x,t) \ge 0.  \label{po}
\end{eqnarray}
Differently from quantum mechanics, however, $\hat{\rho}(t) \neq \int dx \rho(x,t) | x \rangle \langle x |$, 
because, as indicated by Eq. (\ref{2=1}), the pseudo density matrix does not describe the expectation value with 
the so-called mixed state in quantum mechanics.

The above representation of the expectation value with the pseudo density matrix is possible 
for the evolution from the equilibrium state. 
In more general cases, it should be expressed as 
\begin{eqnarray}
\int dx A(x) \rho (x,t)  = \int dx \langle x | \hat{A} | \rho(t) \rangle. \nonumber 
\end{eqnarray}

\section{Irreversible work} \label{sec:work}

Our system is in equilibrium with the control parameter $a_i$ at the initial time $t_i$. 
Now we change this parameter so as to take the value $a_f \equiv a_{t_f}$ at the final time $t_f$.
The work associated with this process is calculated through the expectation value of the change of the potential energy induced by its deformation.
Then the mean work in this process is calculated as \cite{sekimoto-sasa,sekimoto-book,schmiedl-seifert2007}
\begin{eqnarray}
 W 
&=& 
\int^{a_f}_{a_i} da  {\rm Tr} [(\partial_a \hat{V}(a_t)) \hat{\rho}(t)] \nonumber \\
&=& 
\int^{t_f}_{t_i} dt \dot{a}_t  
\sum_n D_n (t) e^{- \theta_n(t) - \int^t_{t_i} ds \bar{\lambda}_n (s) }
\langle 0, a_i | (\partial_a \hat{V} (a_t))| n, a_t \rangle.
\end{eqnarray}
On the other hand, by using Eq.\ (\ref{dn}), the expansion coefficient is iteratively expressed as 
\begin{eqnarray}
D_{n}(t) = 
\delta_{n,0}
- \int^t_{t_i}  dt' 
e^{\int^{t'}_{t_i} ds \bar{\lambda}_n (s) } 
e^{\theta_n(t')}
\frac{\langle n, a_{i} | (\partial_a \hat{\cal L}_{t'}) | 0, a_{t'} \rangle}{\bar{\lambda}_n ({t'}) } \dot{a}_{t'} + O(\dot{a}^2_t).
\end{eqnarray}
Therefore the mean work is expanded in terms of $\dot{a}_t$.

The lowest order calculation of the mean work is given by substituting $D_n(t)$ with $\delta_{n,0}$ 
in the above equation, and then we find
\begin{eqnarray}
W 
&\approx& 
\int^{t_f}_{t_i} dt \dot{a}_t  
\sqrt{\frac{Z(a_i)}{Z(a_t)}} 
\langle 0, a_i | (\partial_a \hat{V} (a_t))| 0, a_t \rangle \nonumber \\
&=& F(a_f) - F(a_i) \equiv \Delta F,
\end{eqnarray}
where $F(a_t)$ is the Helmholtz free energy defined by 
\begin{equation}
F(a_t) = -\frac{1}{\beta} \ln Z(a_t).
\end{equation}
In this derivation, we used 
\begin{eqnarray}
\langle 0, a_i | = \sqrt{\frac{Z(a_t)}{Z(a_i)}} \langle 0, a_t |. \label{eqn:change}
\end{eqnarray}
Note that this simple relation is not satisfied for the ket vector, $| 0, a_i \rangle \neq \sqrt{\frac{Z(a_t)}{Z(a_i)}} | 0, a_t \rangle$.

This result corresponds to the work in the quasi-static process which is given by 
the difference of the Helmholtz free energy.
Thus the irreversible work appears from the higher order terms in $D_n(t)$.

Calculating $D_n(t)$ up to the first order of $\dot{a}_t$, the irreversible work is expressed as 
\begin{eqnarray}
W_{irr}
&=&  W  - \Delta F \nonumber \\ 
&=& 
\int^{t_f}_{t_i}  dt 
\int^t_{t_i} dt'\ \dot{a}_{t'} \Lambda(t',t) \dot{a}_{t} + O(\dot{a}^3_t), \label{ourformula}
\end{eqnarray}
where
\begin{equation}
\Lambda(t',t) = 
-\sum_{n\neq 0}e^{\int^{t'}_{t} ds \bar{\lambda}_n (s) } 
e^{\theta_n(t')-\theta_n(t)}
\frac{\langle n, a_{t'} | (\partial_a \hat{\cal L}_{t'}) | 0, a_{t'} \rangle 
\langle 0, a_{t'} | (\partial_a \hat{V} (a_t))| n, a_t \rangle }{\bar{\lambda}_n (t)}.
\label{ourformula2}
\end{equation}

A similar but different formula to calculate the irreversible work is proposed in Refs.\  \cite{sekimoto-book,sekimoto-sasa}, where 
the solution of the Fokker-Planck equation is expanded with the method which reminds us of the Chapman-Enskog expansion used for the Boltzmann equation.
To reproduce the same result as Refs.\  \cite{sekimoto-book,sekimoto-sasa}, 
we need to ignore the off-diagonal contributions 
in $\Lambda (t',t)$, and 
replace the matrix element calculated from the derivative of the time-dependent Fokker-Planck operator 
($-\langle n, a_{t} | (\partial_a \hat{\cal L}_t) | 0, a_{t} \rangle$)
with that of the potential $(\langle n, a_{t} | (\partial_a \hat{V}(a_t) ) | 0, a_{t} \rangle)$ in Eq.\ (\ref{ourformula2}).
Nevertheless, as is shown in Sec.\ \ref{sec:largetau}, 
our formula gives the same result as Ref.\ \cite{sekimoto-sasa} 
when it is applied to the harmonic potential in the quasi-static limit.

\subsection{Sum rule}

Besides Eqs.\ (\ref{trace}), (\ref{2=1}) and (\ref{po}), our pseudo density matrix $\hat{\rho}(t)$ satisfies another mathematical relation.
When we consider the situation where the system reaches another equilibrium state at $t=t_f$ with $a_f$. 
Because then $D_m (t_f) = 0$ for $m\neq 0$ by definition, we find the following sum rule,   
\begin{equation}
\langle e^{- \beta \int^{t_i}_{t_f} ds \dot{a}_s \partial_a \hat{V}(a_s)} \rangle = e^{-\beta \{ F(a_i) - F(a_f)\}}.
\end{equation}
In this derivation, we used the following mathematical property, 
\begin{eqnarray}
\langle 0, a_i | e^{\hat{G}(a_f) - \hat{G}(a_i)} | m, a_{f} \rangle
= 
 \langle m, a_f | 0, a_i \rangle.
\nonumber 
\end{eqnarray}

It should be noticed that this relation is not the Jarzynski equality itself \cite{jar}. 
As a matter of fact, the above expectation value $\langle~~\rangle$ is not the probability distribution of the work.
However, if we consider a very short time evolution, the change of the particle position will 
be negligibly small and the expectation value with the work distribution might be identified with that 
of the (initial) distribution of the particle. 
Then, the above sum rule can be regarded as a special case of the Jarzynski equality.

\section{Application to harmonic potential} \label{sec:harmonic}

The irreversible work can be calculated exactly for the case of the harmonic potential as is shown 
in Ref.\ \cite{schmiedl-seifert2007}. 
To see the consistency of our expansion, we apply our result to this case.

\subsection{Eigenfunctions} \label{sec:harmo}

Then the time-dependent Fokker-Planck operator is given by 
\begin{eqnarray}
{\cal L}_t(x) = \frac{1}{\nu\beta}\partial^2_x + \frac{a_t}{\nu} \partial_x x.
\end{eqnarray}
This can be transformed as 
\begin{eqnarray}
e^{G(x,a_t)/2}{\cal L}_t (x)e^{-G(x,a_t)/2} = -{\cal H}_t, \nonumber 
\end{eqnarray}
where
\begin{eqnarray}
{\cal H}_t &=& - \frac{1}{\nu\beta} \partial^2_x - \frac{a_t}{2\nu} + \frac{\beta}{4\nu}a^2_t x^2, \nonumber \\
{G}(x,a_t) &=& \frac{\beta}{2}a_t {x}^2. \nonumber
\end{eqnarray}

We introduce the lowering and raising operators,
\begin{eqnarray}
A &=& \frac{1}{\sqrt{a_t \beta}} B = \left( \frac{1}{\sqrt{a_t \beta}}\partial_x + \frac{\sqrt{a_t \beta}}{2} x \right),
\nonumber  \\
A^\dagger &=& \frac{1}{\sqrt{a_t \beta}} B^\dagger = \left( -\frac{1}{\sqrt{a_t \beta}}\partial_x + \frac{\sqrt{a_t \beta}}{2} x \right), \nonumber 
\end{eqnarray}
and then the Hamiltonian operator is expressed as 
\begin{eqnarray}
{\cal H}_t = \frac{a_t}{\nu} A^\dagger A. \nonumber 
\end{eqnarray}
One can easily confirm that the following commutation relations are satisfied, 
\begin{eqnarray}
[A, A^\dagger] = 1, \ \ \ [{\cal H}_t, A^\dagger] = \frac{a_t}{\nu}A^\dagger, \ \ \  
[{\cal H}_t, A] = -\frac{a_t}{\nu}A. \nonumber 
\end{eqnarray}
That is, the eigenfunctions are constructed by operating the raising operators to the ground state 
as is the case of quantum mechanics. 
The eigenvalue of ${\cal H}_t$ is given by $\lambda_n (t)= a_t n /\nu$, and the corresponding eigenfunction is expressed using the Hermite polynomials as 
\begin{eqnarray}
u_n (x,a_t) = \frac{1}{\sqrt{n!}} (A^\dagger)^n u_0 (x,a_t) 
= \sqrt{\frac{1}{2^n n!}\sqrt{\frac{\beta a_t}{2\pi}}} e^{-\beta a_t x^2/4} H_n (\sqrt{\beta a_t/2}x).\nonumber 
\end{eqnarray}
For the properties of the Hermite polynomials $H_n (x)$, see Appendix \ref{app:formulas}.
Therefore, from Eqs.\ (\ref{rho-u1}) and (\ref{rho-u2}), 
the bi-orthogonal system is constructed as 
\begin{eqnarray}
\rho_n (x,a_t) &=& \sqrt{\frac{1}{2^n n!}\sqrt{\frac{\beta a_t}{2\pi}}} e^{-\beta a_t x^2/2} H_n (\sqrt{\beta a_t/2}x),\\
\tilde{\rho}_n (x,a_t) &=& \sqrt{\frac{1}{2^n n!}\sqrt{\frac{\beta a_t}{2\pi}}} H_n (\sqrt{\beta a_t/2}x),
\end{eqnarray}
with the eigenvalue, 
\begin{equation}
\bar{\lambda}_n(t) = \lambda_n(t) = \frac{a_t}{\nu}n.
\end{equation}

\subsection{Irreversible work}

Substituting the above bi-orthogonal system to our formula of the irreversible work (\ref{ourformula}), 
we obtain 
\begin{eqnarray}
W_{irr}
=
\int^{t_f}_{t_i} dt   
\int^t_{t_i} dt'
\dot{a}_{t'} \left[e^{-\int^t_{t'} ds\frac{2a_s}{\nu}}
  \frac{ 1}{2\beta a^2_{t'}} \right] \dot{a}_{t}. \label{har-work}
\end{eqnarray}
In this derivation, we used the following relations, 
\begin{eqnarray}
\langle m, a_t | \dot{\hat{\cal L}}_t | n, a_t \rangle 
&=& 
-\frac{\dot{a}_t}{\nu} \left\{ n \delta_{m,n} + \sqrt{(n+1)(n+2)}\delta_{m,n+2} \right\} , \label{formula1}\\
\theta_n (t) &=& \int^t_{t_i} ds\langle n, a_s | n, \dot{a}_s \rangle 
= 
- \frac{1}{2}\left( n + \frac{1}{2} \right) \ln \frac{a_t}{a_i} , \\
\langle 0, a_i | \frac{1}{2}\hat{x}^2 | n, a_t \rangle 
&=& 
\frac{1}{2} \frac{1}{\beta a_t} \left( \frac{a_i}{a_t} \right)^{1/4} (\sqrt{2}\delta_{n,2} + \delta_{n,0}). \label{0n}
\end{eqnarray}
The derivations are shown in Appendix \ref{app:formulas}.

For the harmonic oscillator, 
higher order corrections do not contribute and Eq.\ (\ref{har-work}) gives the exact expression of the irreversible work. 
Because of Eq.\ (\ref{0n}), the coefficient which can contribute to the calculation of $W_{irr}$ is 
only $D_2(t)$. 
The $N$-th order correction term in the calculation of $D_2(t)$ contains the following product of the matrix elements, 
\begin{eqnarray}
\sum_{n_1 \neq 2} \sum_{n_2 \neq n_1} \cdots \sum_{n_N \neq 0}\langle 2, a_{t_1}|\dot{\hat{\cal L}}_{t_1} 
|n_1,a_{t_1}\rangle \langle n_1, a_{t_2}|\dot{\hat{\cal L}}_{t_2}
|n_2,a_{t_2}\rangle \cdots \langle n_N, a_{t_N}|\dot{\hat{\cal L}}_{t_N}
|0,a_{t_N}\rangle. \nonumber 
\end{eqnarray}
On the other hand, from Eq.\ (\ref{formula1}), the matrix 
$\langle m, a_t |\dot{\hat{\cal L}}_{t} |n,a_t \rangle$ 
has a finite contribution only when $m = n+2$ is satisfied, because $n = m$ is excluded in the sum. 
Therefore the contributions which are higher order than $N = 1$ vanish.

\subsection{Comparison with results from other works}

\subsubsection{irreversible work for harmonic potential}

The exact irreversible work is calculated in Refs.\ \cite{schmiedl-seifert2007,schmiedl-seifert2008, gomez} with the moment method, 
where the variance of the Brownian particle is introduced as  
\begin{eqnarray}
\omega(t) = \langle \hat{x}^2 \rangle = {\rm Tr} [\hat{x}^2 \hat{\rho}(t)], \nonumber
\end{eqnarray}
and then the mean work is expressed as 
\begin{eqnarray}
W = \frac{1}{2} \int^{t_f}_{t_i} ds\ \dot{a}_t \omega(t). \nonumber 
\end{eqnarray}
This is equivalent to our result as is shown below.

Note that, by using the partial integration formula, Eq.\ (\ref{har-work}) can be reexpressed as 
\begin{eqnarray}
W_{irr} 
&=& 
- \Delta F + \frac{1}{2\beta} \int^{t_f}_{t_i} dt \left[ \frac{\dot{a}_t}{a_i}e^{-\frac{2}{\nu}\int^t_{t_i}ds a_s} + \frac{2}{\nu}\dot{a}_t \int^t_{t_i} dt' 
e^{-\frac{2}{\nu}\int^t_{t'}ds a_s}\right]. \nonumber
\end{eqnarray}
On the other hand, as is shown in Ref.\ \cite{schmiedl-seifert2007}, 
the variance satisfies the differential equation, 
\begin{eqnarray}
\partial_t \omega = -\frac{2a_t}{\nu} \omega + \frac{2}{\nu \beta}. \label{seifert-de}
\end{eqnarray}
In short, solving this with the equilibrium initial condition where $\omega(t_i) = 1/(\beta a_i)$, Eq.\ (\ref{har-work}) can be cast into 
the following form, 
\begin{eqnarray}
W_{irr} + \Delta F = \frac{1}{2} \int^{t_f}_{t_i} ds\ \dot{a}_t \omega(t). \nonumber
\end{eqnarray}

\subsubsection{Small and large time-scale limits of external operation $\tau_{op}$} \label{sec:largetau}

Let us investigate the behavior of the irreversible work when the control parameter $a_t$ is changed slowly.
To clarify this limit, we introduce the adimensional time variable \cite{sekimoto-book,sekimoto-sasa} as
\begin{equation}
\tau = \frac{t-t_i}{\tau_{op}} \ \ \ \ \ (0\le \tau \le 1),
\end{equation}
where $\tau_{op} = t_f - t_i$ characterizes the time scale of the external operation.
Then Eq.\ (\ref{har-work}) is reexpressed as 
\begin{equation}
W_{irr}
=
\int^{1}_{0} d\tau   
\int^\tau_{0} d\tau'
\dot{\bar{a}}_{\tau'} \left[e^{- \frac{2\tau_{op}}{\nu}\int^\tau_{\tau'} ds \bar{a}_s}
  \frac{1}{2\beta \bar{a}^2_{\tau'}} \right] \dot{\bar{a}}_{\tau}, \label{exact-har-work}
\end{equation}
where $a_t$ is represented as a function of $\tau$, $\bar{a}_\tau = a_{\tau \tau_{op} + t_i}$.

For the consistency check of the later numerical calculations, it should be noted that, 
in the limit of the instantaneous jump where $\tau_{op} \rightarrow 0$, the irreversible work is simply given by 
\begin{eqnarray}
\lim_{\tau_{op} \rightarrow 0} \frac{W_{irr}}{|\Delta F|} 
= \frac{\left( \frac{\bar{a}_1}{\bar{a}_0} -1 \right)}{|\ln \frac{\bar{a}_1}{\bar{a}_0}|}. \label{work-0-limit}
\end{eqnarray}
This is the exact result independent of the choice of the protocol $\bar{a}_{\tau}$.

In the large limit of $\tau_{op}$, 
the integral for $\tau'$ has the dominant contribution from $\tau' \sim \tau$ because of the exponential factor, 
and thus Eq.\ (\ref{exact-har-work}) is approximately given by 
\begin{equation}
W_{irr}
\approx
\frac{\nu}{4\beta \tau_{op}} \int^{1}_{0} d\tau   
\frac{ \dot{\bar{a}}_{\tau}^2}{\bar{a}^3_{\tau}}. \label{harmonicwork-sekimoto}
\end{equation}
Here we used the approximation which is justified for the large $\tau_{op}$ limit, 
\begin{eqnarray}
\int^\tau_{0} d\tau'
\dot{\bar{a}}_{\tau'} \left[e^{- \frac{2\tau_{op}}{\nu}\int^\tau_{\tau'} ds \bar{a}_s}
  \frac{1}{2\beta \bar{a}^2_{\tau'}} \right] \dot{\bar{a}}_{\tau} 
\approx 
 \frac{\dot{\bar{a}}_{\tau}^2}{2\beta \bar{a}^2_{\tau}}
\int^\tau_{0} d\tau'
e^{- \frac{2\tau_{op}}{\nu}\bar{a}_\tau (\tau- \tau')} 
\approx
\frac{\nu \dot{\bar{a}}_{\tau}^2}{4\beta \tau_{op}\bar{a}^3_{\tau}}. \nonumber
\end{eqnarray}
This approximated expression for the irreversible work is the same as that in Ref.\ \cite{sekimoto-sasa}, which is obtained by 
using different formula for the irreversible work.

The above result (\ref{harmonicwork-sekimoto}) means that the 
irreversible work is proportional to $\tau_{op}^{-1}$ for the large $\tau_{op}$.
It is worth emphasizing that this prediction is experimentally verified. See Ref.\ \cite{blickle2} for details.

\subsection{Optimization}

\begin{figure}[h]
\includegraphics[scale=0.3]{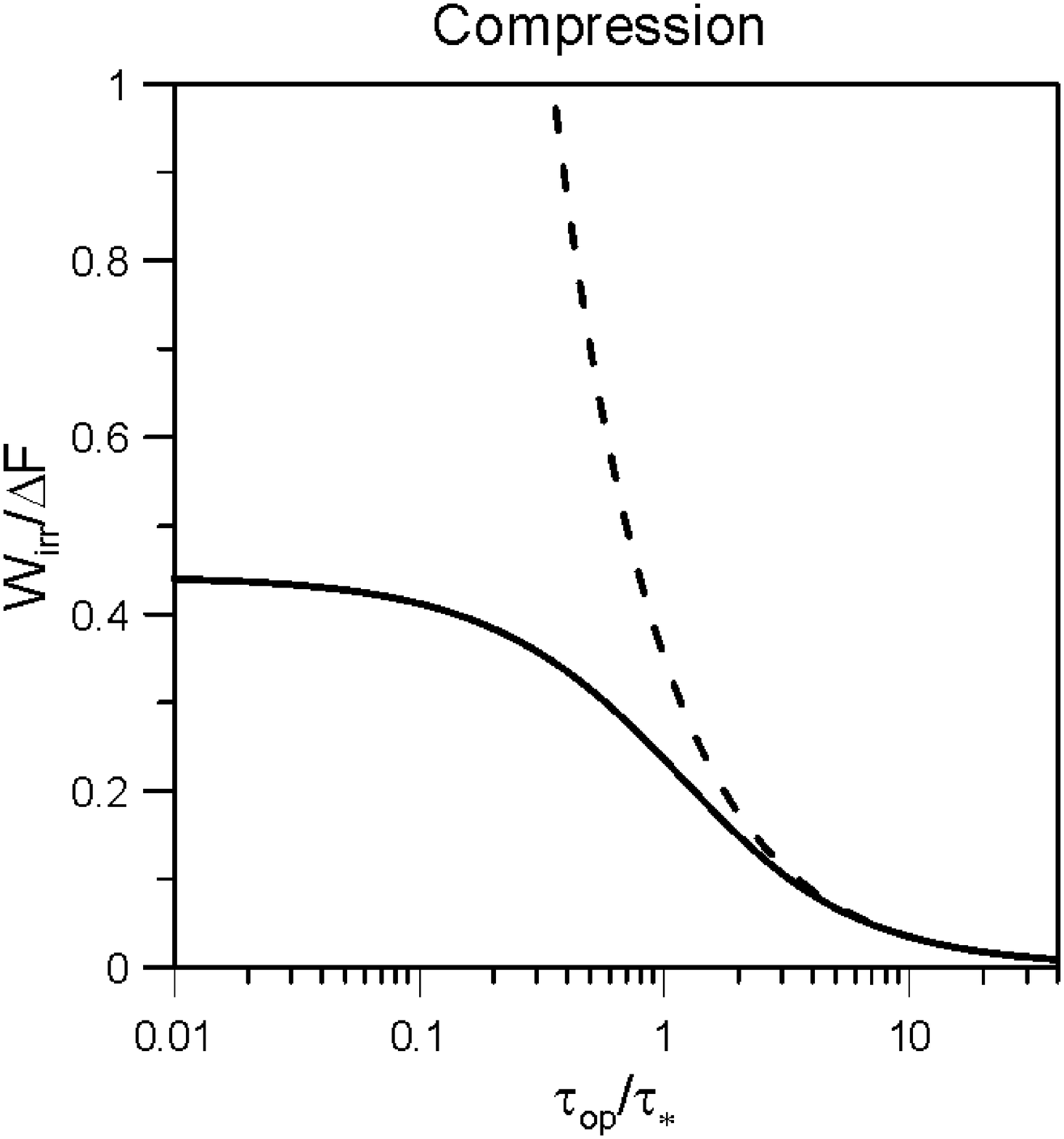}
\includegraphics[scale=0.3]{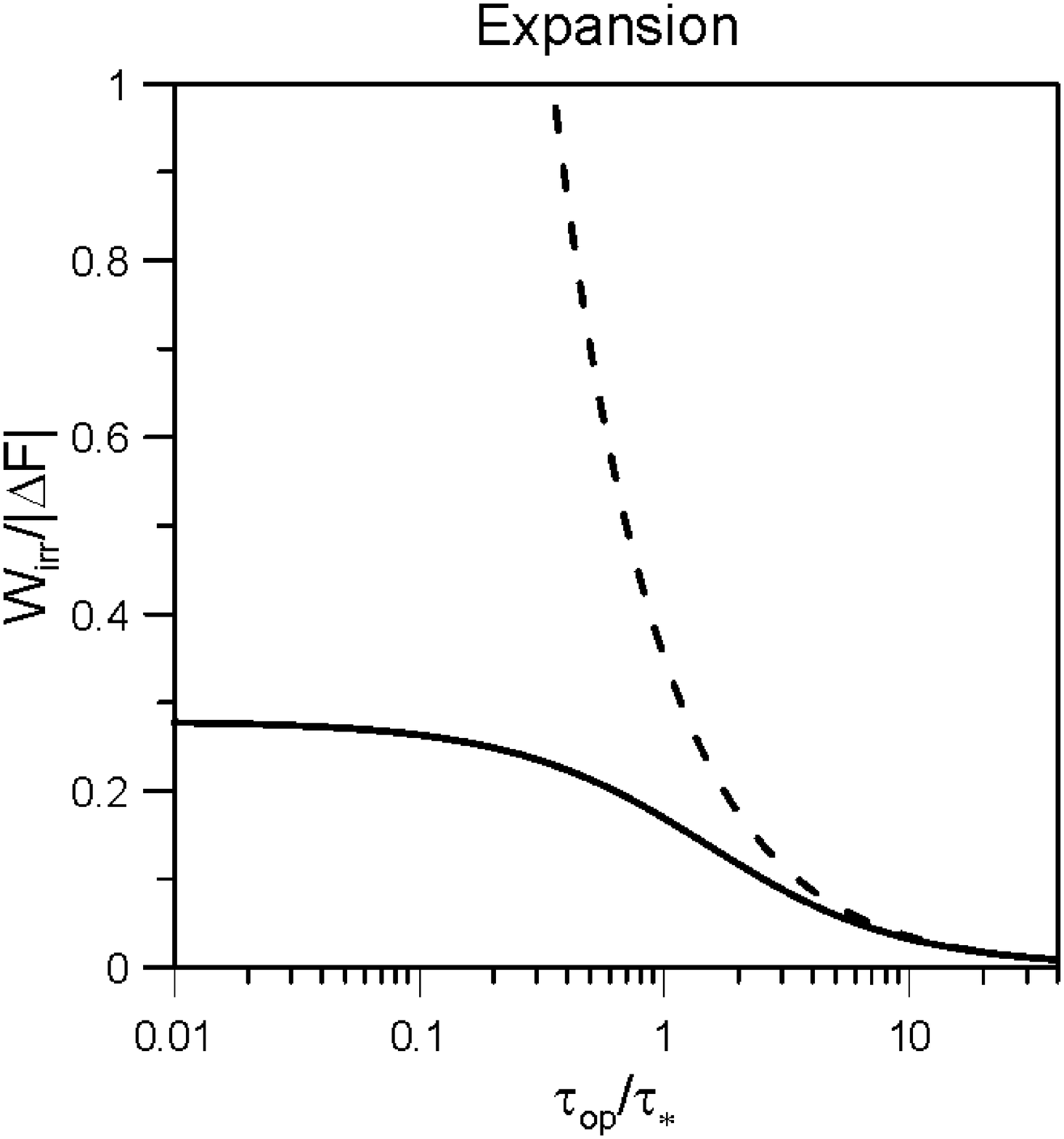}
\caption{The irreversible works are calculated using the control parameter (\ref{parameter-large-t}).
The left and right panels represent the compression process $\bar{a}_1/\bar{a}_0 = 2$ 
and the expansion process $\bar{a}_1/\bar{a}_0 = 1/2$, respectively.
The solid and dashed lines indicate the exact result by Eq.\ (\ref{exact-har-work}) and the approximated one 
by Eq.\ (\ref{harmonicwork-sekimoto}), 
respectively.
We defined ${\tau}_* = \nu/\sqrt{\bar{a}_0 \bar{a}_1}$.}
\label{fig:neq-work1}
\end{figure}

In the Fokker-Planck equation, we can show that the mean work is always larger than $\Delta F$ \cite{sekimoto-book}, 
\begin{eqnarray}
W = \int^{t_f}_{t_i} dt {\rm Tr}[ (\partial_t \hat{V}(a_t)) \hat{\rho}(t) ] \ge \Delta F. \nonumber
\end{eqnarray}
Therefore there exists an optimized control protocol which minimizes the irreversible work for a given time period $\tau_{op}$.

To find the optimized protocol, we calculate the variation of the irreversible work for the following change 
of $a_t$, 
\begin{eqnarray}
a_t \longrightarrow a_t + \delta a_t, \nonumber
\end{eqnarray}
with the fixed initial and final values, 
\begin{eqnarray}
\delta a_i = \delta a_f = 0. \nonumber
\end{eqnarray}
After some of algebra, we find 
\begin{equation}
\int^t_{t_i} ds\frac{\dot{a}_s}{a^2_s} \partial_t e^{-\frac{2}{\nu}\int^t_s d\tau a_\tau}
+ \frac{1}{a^2_t} \int^{t_f}_t ds \dot{a}_s \partial_t e^{-\frac{2}{\nu}\int^s_t d\tau a_\tau} 
+ \frac{2}{\nu} \int^{t_f}_t ds_2 \int^t_{t_i} ds_1 e^{-\frac{2}{\nu}\int^{s_2}_{s_1} d\tau a_\tau} 
\frac{\dot{a}_{s_1}\dot{a}_{s_2}}{a^2_{s_1}} = 0. \label{optimization1}
\end{equation}
See Appendix \ref{app:variation} for more details.

It is difficult to determine the optimized protocol by solving this equation exactly. 
To find an approximated solution, we reexpress this as 
\begin{eqnarray}
&& \bar{a}_\tau  \int^\tau_{0} d\tau_1 g(\tau,\tau_1) \partial_{\tau_1} \frac{\dot{\bar{a}}_{\tau_1}}{\bar{a}^3_{\tau_1}} 
+ \frac{1}{\bar{a}_\tau} \int^{1}_\tau d\tau_1 g(\tau_1,\tau) \partial_{\tau_1} \frac{\dot{\bar{a}}_{\tau_1}}{\bar{a}_{\tau_1}} 
+  \frac{2\tau_{op}}{\nu} \int^{1}_\tau d\tau_2 \int^\tau_{0} d\tau_1 g(\tau_2,\tau_1) \frac{\dot{\bar{a}}_{\tau_1}\dot{\bar{a}}_{\tau_2}}{\bar{a}^2_{\tau_1}} 
\nonumber \\
&& = -\bar{a}_\tau \frac{\dot{\bar{a}}_0}{\bar{a}^3_0} g(\tau,0) + \frac{1}{\bar{a}_\tau} \frac{\dot{\bar{a}}_1}{\bar{a}_1} g(1,\tau),
\label{optimization1D}
\end{eqnarray}
where
\begin{equation}
g(t,s) = e^{-\frac{2\tau_{op}}{\nu}\int^t_s d\tau \bar{a}_\tau}.
\end{equation}
Taking the large $\tau_{op}$ limit, we can employ the same approximation used to obtain the approximated irreversible work (\ref{harmonicwork-sekimoto}). 
Then, for $t_i < t < t_f$, Eq.\ (\ref{optimization1D}) is reduced to  
\begin{eqnarray}
\frac{\ddot{\bar{a}}_t}{\bar{a}^3_t} - \frac{3}{2} \frac{\dot{\bar{a}}^2_t}{\bar{a}^4_t} = 0, \label{eq-opt-app}
\end{eqnarray}
and, near the initial and final times $t \sim t_i, t_f$, we obtain 
\begin{eqnarray}
\dot{\bar{a}}_t = 0.
\end{eqnarray}
The solution of Eq.\ (\ref{eq-opt-app}) is 
\begin{equation}
\bar{a}_\tau = \frac{1}{\{ \bar{a}^{-1/2}_0 - (\bar{a}^{-1/2}_0 - \bar{a}^{-1/2}_1)\tau \}^2} \longrightarrow a_t = \frac{(t_f - t_i)^2 a_i a_f}{\{ (t-t_i) \sqrt{a_i} + (t_f - t) \sqrt{a_f} \}^2}, \label{parameter-large-t}
\end{equation}
which is the same result obtained in Refs.\ \cite{sekimoto-sasa,schmiedl-seifert2007} for the large $\tau_{op}$ limit.

In Fig.\ \ref{fig:neq-work1}, we plotted the irreversible works calculated using the control parameter given 
by Eq.\ (\ref{parameter-large-t}). On the right and left panels, we consider the compression process choosing $\bar{a}_1 = 2 \bar{a}_0$ and 
the expansion process choosing $\bar{a}_0 = 2 \bar{a}_1$, respectively.
We introduced ${\tau}_* = \nu/\sqrt{\bar{a}_0 \bar{a}_1}$ to express the results in adimensional quantities.
To see the applicability of our approximation, the two different irreversible works 
are plotted. The solid line represents the exact result (\ref{exact-har-work}), while the dashed line is 
the approximated one (\ref{harmonicwork-sekimoto}), which is expressed as 
\begin{eqnarray}
\lim_{\tau_{op}\rightarrow 0}\frac{W_{irr}}{\Delta F} = \frac{\sqrt{\bar{a}_0 \bar{a}_1}}{\ln \sqrt{\bar{a}_1/\bar{a}_0}}
\left( \frac{1}{\sqrt{\bar{a}_0}} - \frac{1}{\sqrt{\bar{a}}_1} \right)^2\frac{\tau_*}{\tau_{op}}.
\end{eqnarray}

One can observe that both irreversible works decrease as $\tau_{op}$ increases, because the large $\tau_{op}$ corresponds 
to the quasi-static limit and the contribution from the irreversible work disappears.
For $\tau_{op} \gtrsim \tau_{*}$, our approximation is in good agreement with the exact one, while 
the approximation overestimates the irreversible work for the smaller $\tau_{op}$. 
In particular, the exact result shows that the magnitude of the irreversible work has a finite upper bound and 
does not diverge even in the limit of the instantaneous jump, $\tau_{op} \longrightarrow 0$.
The same behavior is found also in Ref.\ \cite{schmiedl-seifert2007}.
The above mentioned upper bound is given by Eq.\ (\ref{work-0-limit}).

In Ref.\ \cite{schmiedl-seifert2007}, the optimized protocol is obtained by calculating the variation of the irreversible work 
for the variance, not for the control parameter.
The influence of the change of the variational variables is discussed in Sec.\ \ref{sec:conc}.

\subsection{Fluctuation of irreversible work per unit $\tau$}

\begin{figure}[h]
\includegraphics[scale=0.3]{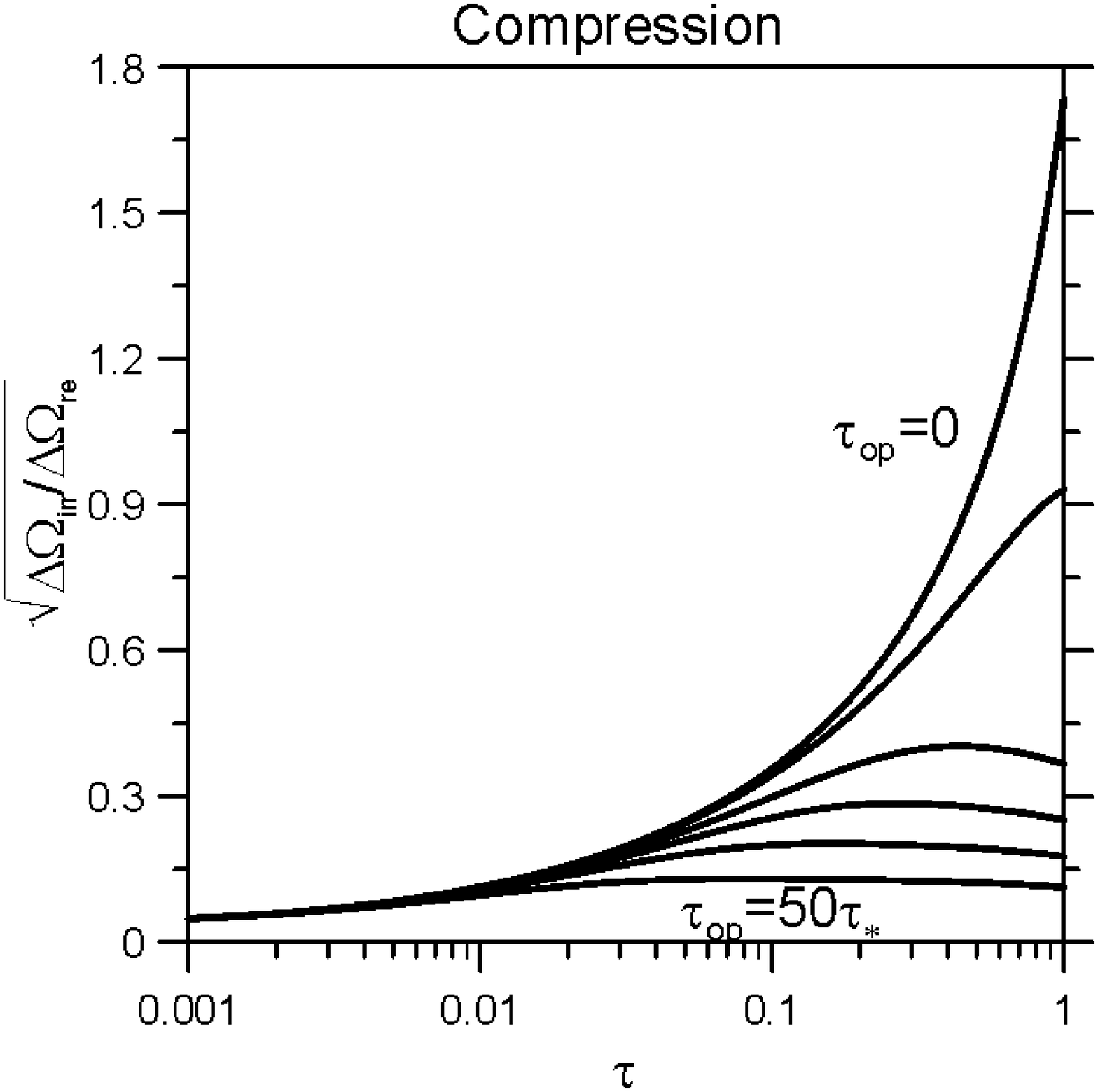}
\includegraphics[scale=0.3]{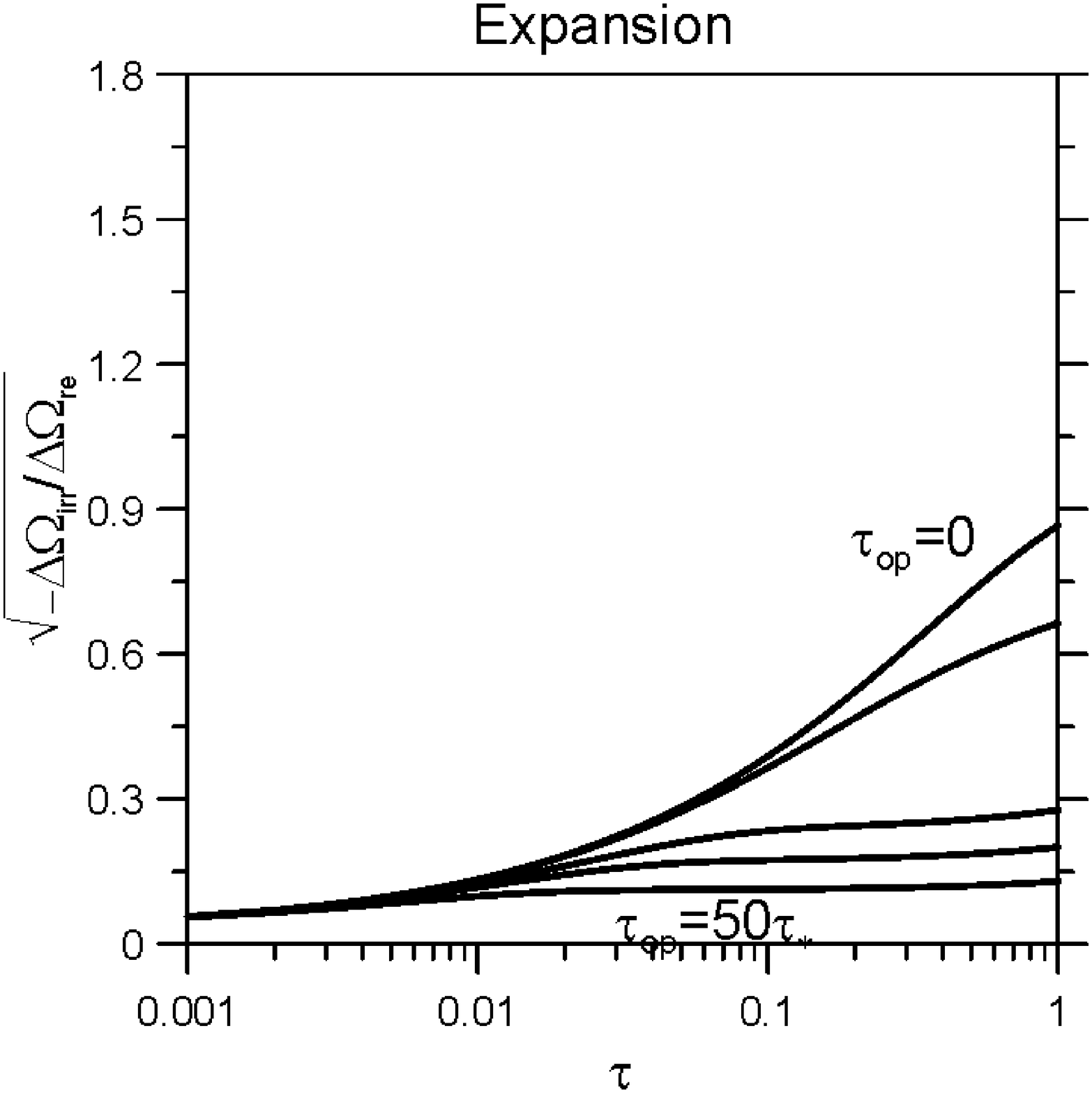}
\caption{The fluctuations of the irreversible work per unit $\tau$ are calculated using the control parameter (\ref{parameter-large-t}).
The left and right panels represent the compression process $\bar{a}_1/\bar{a}_0 = 2$ 
and the expansion process $\bar{a}_1/\bar{a}_0 = 1/2$, respectively.
Each line represent the result for $\tau_{op}/\tau_*=0$, $1$, $5$, $10$, $20$ and $50$ from the top, respectively.
The result is normalized by $\Delta \Omega_{re}$ which is the fluctuation in the quasi-static process. 
Note that this ratio disappears at $\tau=0$.
}
\label{fig:fluctuation}
\end{figure}

The calculations of other correlation functions are simplified in our approach compared to the moment method
because we do not need to solve the coupled differential equations of the moments.  
As an example, we calculate 
the fluctuation of the irreversible work per unit $\tau$, which is given by 
$\langle (d\hat{V}(\bar{a}_\tau)/d\tau)^2 \rangle$.
Then we define the quantity which characterizes the fluctuation as 
\begin{eqnarray}
\Delta \Omega_{irr}(\tau) 
&=& \left\langle \left(\frac{d\hat{V}(\bar{a}_\tau)}{d\tau}\right)^2 \right\rangle - \Delta \Omega_{re}(\tau)
\nonumber \\
&=& 
\frac{\dot{\bar{a}}^2_\tau}{4} \sum_{n\neq 0} D_n (\tau) e^{-\theta_n (\tau) - \tau_{op} \int^\tau_{0} ds \bar{\lambda}_n (s)} 
\langle 0, \bar{a}_0 | \hat{x}^4| n, \bar{a}_\tau \rangle,
\end{eqnarray}
where $\Delta \Omega_{re}(\tau)$ is the fluctuation in the quasi-static process, which is defined by 
\begin{eqnarray}
\Delta \Omega_{re}(\tau) = \lim_{\tau_{op}\rightarrow \infty}\left\langle \left(\frac{d\hat{V}(a_t)}{d\tau}\right)^2 \right\rangle 
= 
\frac{\dot{\bar{a}}^2_\tau}{4} e^{-\theta_0 (\tau)} 
\langle 0, \bar{a}_0 | \hat{x}^4| 0, \bar{a}_\tau \rangle.
\end{eqnarray}

This fluctuation can be calculated exactly for the harmonic oscillator.
From the structure of the matrix element $\langle 0, \bar{a}_0 | \hat{x}^4| 0, \bar{a}_\tau \rangle$, the coefficients which 
contribute to the calculation are only $D_2(\tau)$ and $D_4(\tau)$. 
These are calculated exactly by the first and second order iterations of Eq.\ (\ref{dn}), 
respectively. 
Then we find the exact representation of the fluctuation of the irreversible work per unit $\tau$ as 
\begin{eqnarray}
\Delta \Omega_{irr}(\tau) &=& \frac{3 \sqrt{2}\dot{\bar{a}}^2_\tau}{2\beta^2 \bar{a}_0 \bar{a}_\tau}   D_2 (\tau) 
e^{ - \frac{2\tau_{op}}{\nu} \int^\tau_{0} ds \bar{a}_s} 
+
\frac{\sqrt{6}\dot{\bar{a}}^2_\tau}{2\beta^2 \bar{a}^2_0}  D_4 (\tau) 
e^{- \frac{4\tau_{op}}{\nu} \int^\tau_{0} ds \bar{a}_s }, 
\end{eqnarray}
where
\begin{eqnarray}
D_2 (\tau) &=&
\frac{\bar{a}_0}{\sqrt{2}}\int^\tau_{0} d\tau_1 
e^{\frac{2\tau_{op}}{\nu}\int^{\tau_1}_{0}ds \bar{a}_s} \frac{\dot{\bar{a}}_{\tau_1}}{\bar{a}^2_{\tau_1} }, \\ 
D_4 (\tau) &=& 
\sqrt{3}\bar{a}_0 \int^\tau_{0} d\tau_1 
e^{\frac{2\tau_{op}}{\nu}\int^{\tau_1}_{0} ds \bar{a}_s}
\frac{\dot{\bar{a}}_{\tau_1}}{\bar{a}^2_{\tau_1}} 
 D_2 (\tau_1),
\end{eqnarray}
and
\begin{eqnarray}
\Delta \Omega_{re}(\tau)
=
\frac{3}{4\beta^2}  \frac{\dot{\bar{a}}^2_\tau}{\bar{a}_\tau^2}.
\end{eqnarray}
As is the case of the irreversible work, we can calculate the simple form of this fluctuation in the vanishing limit of $\tau_{op}$ as 
\begin{eqnarray}
\lim_{\tau_{op} \rightarrow 0} \frac{\Delta \Omega_{irr}(\tau)}{\Delta \Omega_{re}(\tau)} = \frac{\bar{a}^2_\tau - \bar{a}^2_0}{\bar{a}^2_0}.
\end{eqnarray}
This is the exact result independent of the choice of $\bar{a}_\tau$.

We consider that the control parameter is again given by Eq.\ (\ref{parameter-large-t}).
Then the numerical results of the fluctuations are shown in Fig.\ \ref{fig:fluctuation}. 
The left and right panels represent the compression process $\bar{a}_1 = 2\bar{a}_0$ 
and the expansion process $\bar{a}_0 = 2\bar{a}_1$, respectively.
Each line represents the result for $\tau_{op}/\tau_*=0$, $1$, $5$, $10$, $20$ and $50$ from the most top line, respectively.
Although it is not plotted, $\Delta \Omega_{irr}/\Delta \Omega_{re}$ disappears at $\tau=0$.
One can see that the magnitudes of the fluctuations do not diverge and take finite values even for the instantaneous jump process.

Note that the time evolutions of the fluctuations are not monotonic and there exist peaks for lines of $\tau_{op} \gtrsim 5 \tau_*$ 
for the compression. 
Such peaks, however, may disappear when we use the optimized parameter by exactly solving Eq.\ (\ref{optimization1}).

\section{Concluding remarks} \label{sec:conc}

In this paper, we developed the framework for the systematic expansion of the solution of the Fokker-Planck equation 
with the help of the eigenfunctions of the time-dependent Fokker-Planck operator. 
The expansion parameter is the time derivative of the external parameter which controls the form of the external confinement potential.
Our expansion corresponds to the perturbative calculation of the adiabatic motion in quantum mechanics. 
With this result, we derived a new formula to calculate irreversible work order by order, 
which is expressed as the expectation value with the pseudo density matrix which can describe only the expectation values with pure states.

By applying this to the harmonic potential, 
we confirmed that the first order calculation gives the exact irreversible work, which is the consistent result 
with that of the moment method 
 \cite{schmiedl-seifert2007,schmiedl-seifert2008,gomez}.
By taking the large $\tau_{op}$ limit, 
we further verified that our formula reproduces the result of Refs.\ \cite{sekimoto-book,sekimoto-sasa} where 
the solution of the Fokker-Planck equation is expanded with the method which reminds us of the Chapman-Enskog expansion. 
However, the structure of the formula in Refs.\ \cite{sekimoto-book,sekimoto-sasa} is qualitatively different 
from ours and such a coincidence will not be seen when it is applied to other potentials.

Higher perturbative corrections can be calculated systematically in our approach 
and the accuracy of the prediction is improved up to arbitrary order, differently from the formula proposed 
in Refs.\ \cite{sekimoto-book,sekimoto-sasa}. 
Moreover, the calculations of correlation functions are simplified compared to 
the moment method because we do not need to solve the coupled differential equations of the moments.
As an example, we showed the calculation of the fluctuation of the irreversible work per unit $\tau$.
These are advantages of our approach.

Because the irreversible work is expressed as an analytic function of the control parameter $a_t$, 
we can apply the variational procedure to find the optimized protocol which minimizes the irreversible work. 
The derived equation, however, is a complex integro-differential equation and difficult to be solved. 
Instead, we discussed the procedure to find the approximated solution. 
As is shown in Fig.\ \ref{fig:neq-work1}, this approximation reproduces the exact behavior of the irreversible work for the large $\tau_{op}$ region, 
and coincides with the results in Refs. \cite{sekimoto-sasa,schmiedl-seifert2007}.

There is a comment for the optimization of the control parameter. 
In Ref.\ \cite{schmiedl-seifert2007}, the optimized parameter for the case of the harmonic potential is obtained 
by calculating the variation of the irreversible work with respect 
to the variance $\omega(t)$ of the position of the Brownian particle, 
instead of the control parameter $a_t$ itself.
However, optimizations generally depend on the choice of the variational variables 
and we cannot change the variables without justification.
Of course, in analytical mechanics, the variational procedure is known to be independent of the choice of coordinates, 
but it is because the variable transformation in such a case is given by the local function of time.
That is, if $\omega(t)$ can be expressed as a function only of $a_t$, 
the variation for $\omega(t)$ leads to the same result as that for $a_t$. 
In the present case, however, $\omega(t)$ is the solution of the differential equation (\ref{seifert-de}) and 
depends on the hysteresis of $a_t$.
Then the variation for $\omega(t)$ does not necessarily coincide with that for $a_t$.
For example, 
let us consider the modification of the optimized function by adding a term which is given by a function of $\omega(t_f)$.
Clearly the variation of this term with respect to $\omega(t)$ vanishes and the optimization is not affected.
However, the added term induces another contribution in the variation with respect to $a_t$ 
because of the hysteresis of $a_t$ in $\omega(t)$. 
In the present optimization problem, therefore, we should consider the variation for $a_t$, not for $\omega(t)$.

For the sake of simplicity, we introduced the bra-ket notation then 
the Fokker-Planck equation (\ref{eqn:fpeq-general}) is expressed as if it is invariant for the choice of the representation.
Then it might be possible to obtain a master equation by multiplying a discretized complete basis to Eq.\ (\ref{eqn:fpeq-general}), 
instead of $\langle x |$.
Master equations are considered to be important in chemical reactions \cite{sekimoto-book} and 
the corresponding eigenvalue problem for the case of time-periodic perturbations is discussed in Ref.\ \cite{caceres}.

The present approach can be extended to more general potentials. 
Because it is generally difficult to find the analytic expressions of the eigenfunctions in non-linear potentials, 
we need to introduce another expansion to find the analytic forms of the expansion basis in our formula. 
For example, we often consider time-periodic protocols and then it will be useful to apply the method in Ref.\ \cite{shirley} 
where the eigenvalue theory for a time-periodic Hamiltonian operator in quantum mechanics is developed.  
Moreover, when we consider systems in the higher spatial dimension, degenerated eigenvalues appear and then the present 
expansion method should be modified.
These applications are left as future works.

In the present calculation, we have considered that the control parameter $a_t$ is a smooth deterministic function of time. 
However it is more realistic to consider the fluctuation of such a protocol and then we need the optimization with respect to stochastic variables. 
The stochastic variation has been discussed in the formulations of the quantum theory and hydrodynamics, 
but the applicability to the optimization of the irreversible work is still an open problem \cite{svm1,svm2,svm3,svm4}.
It is also interesting whether we can apply the similar argument to relativistic \cite{kk-rel} and quantum systems \cite{qse-koide}.

\vspace*{2cm}

The author acknowledges the
ICE group of the Institute of Physics of URFJ 
for fruitful discussions and comments.
This work is financially supported by Conselho
Nacional de Desenvolvimento Cient\'{i}fico e Tecnol\'{o}gico
(CNPq), project 307516/2015-6.

\appendix

\section{Important formulae for calculations of harmonic potential} \label{app:formulas}

In the following we derive various formulae with the help of the Hermite polynomials which satisfy 
\begin{eqnarray}
H_{n+1}(x) &=& 2x H_n (x) - 2n H_{n-1}(x), \\
\partial_x H_n(x) &=& 2n H_{n-1}(x), \\
\partial_x H_n(x) &=& 2xH_n (x) - H_{n+1}(x), \label{hermite3}\\
\int dx H_m (x) H_n (x) e^{-x^2} &=& 2^n n! \sqrt{\pi} \delta_{n,m}.
\end{eqnarray}

\subsection{$\langle x | \dot{\hat{\cal L}}_{t} | n , a_t \rangle$}

\begin{eqnarray}
\langle x | \dot{\hat{\cal L}}_{t} | n , a_t \rangle 
&=& 
\frac{\dot{a}_t}{\nu} (1 + x \partial_x) \rho_n (x,a_t) \nonumber \\
&=&
\frac{\dot{a}_t}{\nu} c_n e^{-\beta a_t x^2/2}
\left\{ 1 - \beta a_t x^2 + x \partial_x \right\} H_n (\sqrt{\beta a_t/2}x) 
\nonumber \\
&=&
\frac{\dot{a}_t}{\nu} c_n e^{-\beta a_t x^2/2}
\left\{ H_n (\sqrt{\beta a_t/2}x) - \sqrt{\beta a_t/2}x H_{n+1}(\sqrt{\beta a_t/2}x)  \right\} \nonumber \\
&=& 
\frac{\dot{a}_t}{\nu} c_n e^{-\beta a_tx^2/2}
\left\{ H_n (\sqrt{\beta a_t/2}x) 
-\frac{H_{n+2}(\sqrt{\beta a_t/2}x)  + 2(n+1)H_n(\sqrt{\beta a_t/2}x) }{2}  \right\} \nonumber \\
&=& 
-\frac{\dot{a}_t}{\nu} 
\left\{ 
n \langle x | n, a_t \rangle 
+ \frac{1}{2} \frac{c_n}{c_{n+2}} \langle x | n+2, a_t \rangle 
\right\},
\end{eqnarray}
where 
\begin{equation}
c_n = \sqrt{\frac{1}{2^n n!} \sqrt{\frac{\beta a_t}{2\pi}}}.
\end{equation}
From the second to the third line, we used Eq.\ (\ref{hermite3}).
Therefore
\begin{eqnarray}
\langle m, a_t | \dot{\hat{\cal L}}_{t} | n, a_t \rangle 
=
- \frac{\dot{a}_t}{\nu}  \left\{ 
 n \delta_{m,n}
+ \sqrt{(n+1)(n+2)} \delta_{m,n+2}  
\right\}.
\end{eqnarray}

\subsection{$ \langle n, a_t | n, \dot{a}_t \rangle $}

\begin{eqnarray}
\langle n, a_t | n, \dot{a}_t \rangle 
&=& 
\frac{1}{2^n n! \sqrt{\pi}} \int dx 
\sqrt{\frac{\beta}{2\pi}} a^{1/4}_t H_n (\sqrt{\beta a_t /2} x) \partial_t \left( a^{1/4}_t
H_n (\sqrt{\beta a_t /2} x) e^{- \beta a_t /2 x^2}\right) \nonumber \\
&=& 
\frac{1}{2^{n+1} n! \sqrt{\pi}}\frac{\dot{a}_t}{a_t} \int d\xi  H_n (\xi) 
\left\{
\frac{1}{2}  H_n (\xi) + \xi (\partial_\xi H_n (\xi)) - 2\xi^2 H_n (\xi)
\right\} e^{-\xi^2} 
\nonumber \\
&=& 
\frac{1}{2^{n+1} n! \sqrt{\pi}}\frac{\dot{a}_t}{a_t} \int d\xi  H_n (\xi) 
\left\{
\frac{1}{2}  H_n (\xi) + \xi (\partial_\xi H_n (\xi)) - [\xi (\partial_{\xi} H_n (\xi)) + \xi H_{n+1}(\xi)]
\right\} e^{-\xi^2} 
\nonumber \\
&=& 
\frac{1}{2^{n+1} n! \sqrt{\pi}}\frac{\dot{a}_t}{a_t} \int d\xi  H_n (\xi) 
\left\{
\frac{1}{2}  H_n (\xi)  -  \xi H_{n+1}(\xi)
\right\} e^{-\xi^2} 
\nonumber \\
&=& 
\frac{1}{2^{n+1} n! \sqrt{\pi}}\frac{\dot{a}_t}{a_t}  \int d\xi H_n (\xi) 
\left\{
\frac{1}{2}  H_n (\xi)  -  \frac{H_{n+2}(\xi) + 2(n+1)H_n (\xi)}{2}
\right\} e^{-\xi^2} 
\nonumber \\
&=& 
\frac{1}{2^{n+1} n! \sqrt{\pi}} \frac{\dot{a}_t}{a_t} \int d\xi
 H_n (\xi) 
\left\{-\frac{1}{2} H_{n+2}(\xi) - \left( n + \frac{1}{2} \right)H_n (\xi)
\right\} e^{-\xi^2} 
\nonumber \\
&=& 
- \left( n + \frac{1}{2} \right)  \frac{\dot{a}_t}{2a_t},
\end{eqnarray}
where we introduced
\begin{eqnarray}
\xi = \sqrt{\beta a_t /2} x.
\end{eqnarray}

\subsection{$\langle 0, a_i|\frac{1}{2} \hat{x}^2 | n, a_t \rangle $}

\begin{eqnarray}
\langle 0, a_i|\frac{1}{2} \hat{x}^2 | n, a_t \rangle 
&=& 
\frac{1}{2} \int dx \left( \frac{\beta a_i}{2\pi} \right)^{1/4}
x^2 \sqrt{\frac{1}{2^{n} n!}\sqrt{\frac{\beta a_t}{2\pi}}}e^{-\beta a_t x^2/2} H_{n} \left( \sqrt{\beta a_t/2}x \right) 
\nonumber \\
&=& 
\frac{1}{2}\sqrt{\frac{1}{2^{n} n!}\frac{1}{\pi}\sqrt{\frac{a_i}{a_t}}} \frac{2}{\beta a_t}
 \int d\xi 
\xi^2 e^{-\xi^2} H_{n} \left( \xi \right) 
\nonumber \\
&=& 
\sqrt{\frac{1}{2^{n} n!}\frac{1}{\pi}\sqrt{\frac{a_i}{a_t}}} \frac{1}{\beta a_t}
 \int d\xi 
\frac{1}{4} \left\{ H_2 (\xi) + 2 H_0 (\xi)\right\} H_{n} \left( \xi \right)e^{-\xi^2}  
\nonumber \\
&=& 
\frac{1}{2}\frac{1}{\beta a_t} \left( \frac{a_i}{a_t}\right)^{1/4}
( \sqrt{2} \delta_{n,2} + \delta_{0,n} ).
\end{eqnarray}

\section{Variation} \label{app:variation}

The change of the irreversible work for the variation of $a_t$ is calculated as 
\begin{eqnarray}
\delta W &=& W[a_t + \delta a_t] - W[a_t] \nonumber \\
&& =
\frac{1}{2\beta} \int^{t_f}_{t_i} dt \int^{t}_{t_i} ds \frac{g(t,s) \dot{a}_s }{a^2_s} \delta \dot{a}_t 
+ \frac{1}{2\beta}\int^{t_f}_{t_i} dt \int^{t}_{t_i} ds \delta \dot{a}_s \frac{ g(t,s) \dot{a}_t }{a^2_s}\nonumber \\
&& - \frac{1}{\beta} \int^{t_f}_{t_i} dt \int^{t}_{t_i} ds \frac{g(t,s) \dot{a}_t \dot{a}_s }{a^3_s}  \delta a_s
-\frac{1}{\beta \nu} \int^{t_f}_{t_i} dt \int^{t}_{t_i} ds \int^t_s d\tau \frac{g(t,s) \dot{a}_t \dot{a}_s }{a^2_s}  \delta a_\tau  
\nonumber \\
&& = 
- \frac{1}{2\beta} \int^{t_f}_{t_i} dt \delta a_t \left[ \frac{d}{dt} \int^{t}_{t_i} ds \frac{g(t,s) \dot{a}_s }{a^2_s} 
+ \frac{d}{dt} \int^{t_f}_{t} ds \frac{g(s,t) \dot{a}_s }{a^2_t} 
\right] 
- \frac{1}{\beta} \int^{t_f}_{t_i} dt \delta a_t \int^{t_f}_{t} ds \frac{g(s,t) \dot{a}_s \dot{a}_t }{a^3_t} 
\nonumber \\
&& -\frac{1}{\beta \nu} \int^{t_f}_{t_i} dt \int^{t_f}_{t_i} ds \int^{t_f}_{t_i} d\tau \theta(t-s)\theta(t-\tau)\theta(\tau-s) \frac{g(t,s) \dot{a}_t \dot{a}_s }{a^2_s}  \delta a_\tau  \nonumber \\
&& = 
- \frac{1}{2\beta} \int^{t_f}_{t_i} dt \delta a_t \left[ \frac{d}{dt} \int^{t}_{t_i} ds \frac{g(t,s) \dot{a}_s }{a^2_s} 
+ \frac{d}{dt} \int^{t_f}_{t} ds \frac{g(s,t) \dot{a}_s }{a^2_t} 
\right] \nonumber \\
&& - \frac{1}{\beta} \int^{t_f}_{t_i} dt \delta a_t \int^{t_f}_{t} ds \frac{g(s,t) \dot{a}_s \dot{a}_t }{a^3_t} 
-\frac{1}{\beta \nu} \int^{t_f}_{t_i} dt \delta a_t \int^{t_f}_{t} d\tau \int^{t}_{t_i} ds 
\frac{g(\tau,s) \dot{a}_\tau \dot{a}_s }{a^2_s},
\end{eqnarray}
where
\begin{eqnarray}
g(t,s) = e^{-\frac{2}{\nu}\int^t_s d\tau a_\tau}.
\end{eqnarray}
Therefore the optimized protocol is described by the following equation, 
\begin{eqnarray}
&& \frac{d}{dt} \left[ \int^{t}_{t_i} ds \frac{g(t,s) \dot{a}_s }{a^2_s} 
+ \int^{t_f}_{t} ds \frac{g(s,t) \dot{a}_s }{a^2_t} \right] 
+ 2\int^{t_f}_{t} ds \frac{g(s,t) \dot{a}_s \dot{a}_t }{a^3_t} 
+ \frac{2}{\nu}\int^{t_f}_{t} d\tau \int^{t}_{t_i} ds 
\frac{g(\tau,s) \dot{a}_\tau \dot{a}_s }{a^2_s}= 0 \nonumber \\
&& \longrightarrow 
\int^{t}_{t_i} ds \frac{\dot{a}_s }{a^2_s} \partial_t g(t,s) 
+ \int^{t_f}_{t} ds \frac{\dot{a}_s }{a^2_t}\partial_t g(s,t) 
+ \frac{2}{\nu}\int^{t_f}_{t} d\tau \int^{t}_{t_i} ds 
\frac{g(\tau,s) \dot{a}_\tau \dot{a}_s }{a^2_s}= 0.
\end{eqnarray}


\begin{thebibliography}{10}


\bibitem{hanngi-review}
P.\ H\"{a}nggi and F.\ Marchesoni, 
``Artificial Brownian motors: Controlling transport on the nanoscale", Rev. Mod. Phys. \textbf{81}, 387 (2009).

\bibitem{broeck-review}
C.\ Van den Broeck, S.\ Sasa, and U.\ Seifert, 
``Focus on stochastic thermodynamics", New J. Phys. \textbf{18}, 020401 (2016).

\bibitem{sei-rev}
U.\ Seifert,
``Stochastic thermodynamics, fluctuation
theorems and molecular machines", 
Rep. Prog. Phys. \textbf{75}, 126001 (2012).



\bibitem{sekimoto-book}
K.\ Sekimoto, {\it Stochastic Energetics} (Springer, Berlin, 2010).


\bibitem{sekimoto-sasa}
K.\ Sekimoto and S.\ Sasa, 
``Complementarity relation for irreversible process derived from stochastic energetics", 
J. Phys. Soc. Jpn. \textbf{66}, 3658 (2001).


\bibitem{koning}
M.\ de Koning, 
``Optimizing the driving function for nonequilibrium free-energy calculations
in the linear regime: A variational approach", 
J. Chem. Phys. \textbf{122}, 104106 (2005).



\bibitem{all}
A.\ E.\ Allahverdyan and Th.\ M.\ Nieuwenhuizen, 
``Minimal-work principle and its limits for classical systems",
Phys. Rev. E\textbf{75}, 051124 (2007).

\bibitem{then}
H.\ Then and A.\ Engel, 
``Computing the optimal protocol for finite-time processes in stochastic thermodynamics", 
Phys. Rev. \textbf{E77}, 041105 (2008).


\bibitem{geiger}
P.\ Geiger and C.\ Dellago, 
``Optimum protocol for fast-switching free-energy calculations", 
Phys. Rev. \textbf{E81}, 021127 (2010).


\bibitem{aurell}
E.\ Aurell, C.\ Mej\'{i}a-Monasterio and P.\ Muratore-Ginanneschi, 
``Optimal Protocols and Optimal Transport in Stochastic Thermodynamics",
Phys. Rev. Lett. \textbf{106}, 250601 (2011).


\bibitem{crook}
D.\ A.\ Sivak and G.\ E.\ Crook, 
``Thermodynamic Metrics and Optimal Paths", 
Phys. Rev. Lett. \textbf{108}, 190602 (2012).

\bibitem{bona}
M.\ V.\ S.\ Bonan\c{c}a and S.\ Deffer, 
``Optimal driving of isothermal processes close to equilibrium", 
J. Chem. Phys. \textbf{140}, 244119 (2014).


\bibitem{dechant}
A.\ Dechant, N.\ Kiesel and E.\ Lutz, 
``All-Optical Nanomechanical Heat Engine", 
Phys. Rev. Lett. \textbf{114}, 183602 (2015).


\bibitem{zulkow}
P.\ R.\ Zulkowski and M.\ R.\ DeWeese,
``Optimal control of overdamped systems"
Phys. Rev. E\textbf{92}, 032117 (2015).


\bibitem{schmiedl-seifert2007}
T.\ Schmiedl and U.\ Seifert, 
``Optimal Finite-Time Processes In Stochastic Thermodynamics", Phys. Rev. Lett. \textbf{98}, 108301 (2007).

\bibitem{schmiedl-seifert2008}
T.\ Schmiedl and U.\ Seifert, 
``Efficiency at maximum power: An analytically solvable model for stochastic heat engines", EPL \textbf{81}, 20003 (2008).

\bibitem{gomez}
A.\ Gomez-Marin, T.\ Schmiedl and U.\ Seifert,
``Optimal protocols for minimal work processes in underdamped
stochastic thermodynamics", 
J. Chem. Phys. \textbf{129}, 024114 (2008).


\bibitem{speck}
T.\ Speck, 
``Work distribution for the driven harmonic oscillator with time-dependent strength: exact solution and slow driving", 
J. Phys. A: Math. Theor. \textbf{44}, 305001 (2011).

\bibitem{mazon}
O.\ Mazonkaa and C.\ Jarzynski, 
``Exactly solvable model illustrating far-from-equilibrium predictions", 
arXiv:cond-mat/9912121 (1999).

\bibitem{raybov}
A.\ Ryabov, et al., 
``Work distribution in a time-dependent logarithmic-harmonic potential: exact results and
asymptotic analysis", 
J. Phys. A: Math. Theor. \textbf{46}, 075002 (2013).


\bibitem{imparato}
A.\ Imparato, et al., 
``Work and heat probability distribution of an optically driven Brownian particle:
Theory and experiments", 
Phys. Rev. E\textbf{76}, 050101(R) (2007).

\bibitem{cohen}
R.\ van Zon and E.\ G.\ D.\ Cohen, 
``Stationary and transient work-fluctuation theorems for a dragged Brownian particle",
Phys. Rev. E\textbf{67}, 046102 (2003).

\bibitem{kwon}
C.\ Kwon, J.\ D.\ Noh and H.\ Park,
``Work fluctuations in a time-dependent harmonic potential: Rigorous results beyond
the overdamped limit", 
Phys. Rev. E\textbf{88}, 062102 (2013).

\bibitem{hol2}
V.\ Holubec and A.\ Ryabov, 
``Efficiency at and near maximum power of low-dissipation heat engines", 
Phys. Rev. E\textbf{92}, 052125 (2015).

\bibitem{hol}
V.\ Holubec, 
``An exactly solvable model of a stochastic heat engine: optimization of
power, power fuctuations and efficiency", 
J. Stat. Mech. P05022, (2014).


\bibitem{blickle}
V.\ Blickle et al., 
``Thermodynamics of a Colloidal Particle in a Time-Dependent Nonharmonic Potential",
Phys. Rev. Lett. \textbf{96}, 070603 (2006).

\bibitem{dieterich}
T.\ Schmiedl, et al., 
``Optimal protocols for Hamiltonian and Schr\"{o}dinger dynamics", 
J. Stat. Mech. P07013, (2009).


\bibitem{gardiner}
C.\ W.\ Gardiner, {\it Handbook of Stochastic Method: For Physics,
Chemistry and Natural Sciences} (Springer, New York, 2004).


\bibitem{namiki}
We consulted the discussion also in M.\ Namiki, {\it Delta Function and Differential Equation}, (Iwanami, Tokyo, 1982) (in Japanese).

\bibitem{caceres}
M.\ O.\ Cacerea and A.\ M.\ Lobos, 
``Theory of eigenvalues for periodic non-stationary Markov processes: the Kolmogorov operator and
its applications", J. Phys. A: Math. Gen. \textbf{39}, 1547 (2006).


\bibitem{talkner}
P.\ Talkner and J.\ {\L}uczka, 
``Rate description of Fokker-Planck processes with time-dependent parameters", 
Phys. Rev. E\textbf{69}, 046109 (2004).


\bibitem{dreger}
J.\ Dreger, A.\ Pelster and B.\ Hamprecht, 
``Variational Perturbation Theory for Fokker-Planck Equation with Nonlinear Drift", 
Euro. Phys. J. B\textbf{45}, 355 (2005).

\bibitem{hesam}
S.\ Hesam, A.\ R.\ Nazemi and A.\ Haghbin, 
``Analytical solution for the Fokker-Planck equation by differential
transform method", 
Scientia Iranica B \textbf{19}, 1140 (2012).

\bibitem{welles2014}
W.\ A.\ M.\ Morgado and S.\ M.\ Duarte Queir\'{o}s, 
``Thermostatistics of small nonlinear systems: Gaussian thermal bath", 
Phys. Rev. E\textbf{90}, 022110 (2014).


\bibitem{berry}
J.\ C.\ Budich and B.\ Trauzettel,
``From the adiabatic theorem of quantum mechanics to topological states of matter",
Phys. Status Solidi (RRL) \textbf{7} 109 (2013).


\bibitem{jar}
C.\ Jarzynski, 
``Nonequilibrium Equality for Free Energy Differences", Phys. Rev. Lett. \textbf{78}, 2690 (1997).


\bibitem{blickle2}
V.\ Blickle and C.\ Bechinger, 
``Realization of a micrometre-sized stochastic
heat engine", Nat. Phys. \textbf{8}, 143 (2012).


\bibitem{shirley}
J.\ H.\ Shirley, 
``Solution of the Schr\"{o}dinger Equation with a Hamiltonian Periodic in Time", 
Phys. Rev. \textbf{138}, B979 (1965).


\bibitem{svm2}
T.\ Koide and T.\ Kodama, 
``Navier-Stokes, Gross-Pitaevskii and generalized diffusion equations using the
stochastic variational method", J. Phys. A: Math. Theor. \textbf{45}, 255204 (2012).


\bibitem{svm1}
T.\ Koide, T.\ Kodama and K.\ Tsushima, 
``Unified description of classical and quantum behaviours in a variational principle", 
J. Phys.: Conf. Ser. \textbf{626}, 012055 (2015).


\bibitem{svm3}
T.\ Koide and T.\ Kodama,
``Stochastic Variational Method as Quantization Scheme : Field Quantization of Complex Klein-Gordon Equation", 
Prog. Theor. Exp. Phys. 093A03, (2015).


\bibitem{svm4}
T.\ Koide
``Classicalization of quantum variables and quantum-classical hybrids", 
Phy. Lett. A\textbf{379}, 2007 (2015).


\bibitem{kk-rel}
T.\ Koide and T.\ Kodama, 
``Thermodynamic laws and equipartition theorem in relativistic Brownian motion", 
Phys. Rev. E\textbf{83}, 061111 (2011).


\bibitem{qse-koide}
T.\ Koide, 
``Memory effect in the upper bound of the heat flux induced by quantum fluctuations", 
Phys. Rev. E\textbf{94}, 042140 (2016).

\end{thebibliography}
\end{document}